\documentstyle[prd,eqsecnum,aps,psfig]{revtex}

\def\be{\begin{equation}}
\def\ee{\end{equation}}
\def\ba{\begin{eqnarray}}
\def\ea{\end{eqnarray}}

\def\negenspace{\kern-1.1em}

\begin{document}

\input epsf

\title{Boson Stars: Alternatives to primordial black holes?}

\author{Eckehard W. Mielke$^{\diamond}$\thanks{E-mail: ekke@xanum.uam.mx} 
and Franz E.~Schunck$^\$$\thanks{E-mail: fs@astr.cpes.susx.ac.uk}} 
\address{$^{\diamond}$ Departamento de F\'{\i}sica,
Universidad Aut\'onoma Metropolitana--Iztapalapa,\\
Apartado Postal 55-534, C.P. 09340, M\'exico, D.F., MEXICO\\  and\\
$^\$$Astronomy Centre, School of Chemistry, Physics and
Environmental Science,\\ University of Sussex, Falmer, Brighton BN1 9QJ,
United Kingdom,}

\date{\today}

\maketitle

\begin{abstract}
The present surge for the astrophysical relevance of {\em boson stars} stems
from the speculative possibility that these compact objects could provide a
considerable fraction of the
non-baryonic part of {\em dark matter} within the halo of galaxies. 
For a very 
light `universal' {\em axion} of {\em effective string} models, 
their total gravitational mass will be in the most likely range of
$\sim 0.5$ $M_\odot$ of MACHOs. According to this framework, 
{\em gravitational microlensing} is indirectly 
``weighing" the axion mass, resulting in $\sim 10^{-10}$ eV/c$^2$. 
This conclusion is not changing much, if we use a
dilaton type self-interaction 
for the bosons. Moreover,
we review their formation, rotation and stability as likely
candidates of astrophysical importance.
\end{abstract}
\pacs{PACS no.: 04.40.Dg, 04.25.Dm, 95.30.Sf} Keywords: Boson stars, Axions,
Effective string models, Dark matter, MACHOs

\section{Introduction}
\subsection{\bf Dark matter --- Issue of missing mass}

The rotation velocities of spiral galaxies
can be accurately measured from the Doppler
effect. At large radii where the stellar surface brightness is falling
exponentially, velocities are obtained for clouds of 
neutral hydrogen using the 21 cm hyperfine line. 
The resulting `rotation curves' are found to be roughly 
flat out to the maximum observed radii of about 50 kpc,
which implies an enclosed mass increasing linearly with radius. 
This mass profile is much more extended than the distribution 
of starlight, which typically converges within $\sim 10$ kpc; 
thus, the galaxies are presumed to be surrounded by extended 
{\em halos} of {\em dark matter}.   

Perhaps the most compelling evidence for dark matter comes from
clusters of galaxies.
These are structures of about 1 Mpc size 
containing more than 100 galaxies, representing an overdensity of 
about a factor 1000 relative to the mean galaxy density. 
It is assumed that they are gravitationally bound since the
time for galaxies to cross the cluster lasts only about 10\%\ of the
age of the Universe. The cluster masses are determined in several
independent ways: 
First, the virial theorem uses the radial velocities of
individual galaxies as `test particles'.
Second, observations of hot gas at about $10^7$ K
contained in the clusters, which is observed in X-rays
via thermal bremsstrahlung. The gas temperature 
is derived from the X-ray spectrum, and the density profile from
the map of the X-ray surface brightness. 
Assuming the gas is pressure-supported against the gravitational
potential leads to a mass profile for the cluster.  
The third method is gravitational lensing of background objects by the
cluster potential. There are two regimes: the `strong lensing' 
regime at small radii, which leads to arcs and multiple images, and the
weak lensing regime at large radii, which causes background galaxies to be 
preferentially stretched in the tangential direction. 
All three methods yield estimates for cluster masses
\cite{C98,BN92} which show that
visible stars contribute only a few percent of the observed mass,
and the hot X-ray gas only about 10--20\%, hence, clusters
are dominated by dark matter.   

On the largest scales, there is further evidence for dark matter: 
`streaming motions' of galaxies (e.g., towards nearby 
superclusters such as the ``Great Attractor'')
can be compared to maps of the galaxy density from redshift surveys
\cite{pecvel} to yield estimates of $\Omega$.  
Here the theory is more
straightforward since the density perturbations are still in the linear
regime, but the observations are less secure. 
A similar estimate may be derived by comparing our Galaxy's 
600 km/s motion towards the Virgo cluster relative to the cosmic rest
frame, confirmed by the observed temperature
dipole in the cosmic microwave background (CMB).  

Furthermore, it is possible to connect the observed 
large-scale structure in the galaxy distribution with 
the results of the CMB anisotropies if the
universe is dominated by non-baryonic dark matter. Commonly, the present 
matter/energy density $\Omega_0= \Omega_{\rm M} + \Omega_\Lambda$ 
of the Universe is decomposed into two components. 
There is accumulating evidence for 
$\Omega_0 = 1 \pm 0.2$ and (total) 
matter density $\Omega_{\rm M}= 0.4 \pm 0.1$ which implies 
a vacuum energy or {\em dimensionless cosmological constant} 
of  $\Omega_\Lambda= 0.85 \pm 0.2$ \cite{Perlmutter98,riess98},
cf. \cite{Tu99}.
Theories of  inflation prefer a {\em flat} Universe with 
 $\Omega_0 = 1$ as its
most `natural' value; this  also  requires {\em non-baryonic} 
dark matter.

\subsection{\bf Dark matter --- Candidates}

What are realistic candidates for dark matter?
Hot gas appears to be excluded \cite{carr} by limits on 
the Compton distortion of the blackbody CMB spectrum; 
atomic hydrogen due to 21 cm observations; and
ordinary stars by faint star counts. 
Asteroids are very unlikely since stars do not process hydrogen into
heavy elements very efficiently and
hydrogen `snowballs' should evaporate or lead to excessive cratering
on the Moon. Black holes more massive than 
$\sim 10^5$ $M_\odot$ would destroy small globular clusters by tidal effects.
 
Today's most viable dark matter candidates fall into two broad classes: 
astrophysical size objects called {\em MA}ssive {\em C}ompact {\em H}alo 
{\em O}bjects (MACHOs), and so-called {\em W}eakly {\em I}nteracting 
{\em M}assive {\em P}articles (WIMPs). Actually these classes could possibly
be interrelated, as we are going to show. 

Several different objects belong to the first class: 
Jupiter-size brown dwarfs consisting  of hydrogen and helium less
massive than 0.08 $M_\odot$ are the most prominent possibility.
Below this limit, the central temperature is not sufficient in order 
to ignite hydrogen fusion, so these objects just
radiate very weakly in the infrared due to gravitational contraction.
Other MACHO candidates include stellar remnants such as cool
white dwarfs, neutron stars, and primordial black holes or as a result
of a supernova. 

The WIMP candidates are
the invisible axion (hypothesized to solve the strong CP problem or reemerging 
`universally' in {\em effective string} Lagrangians),
one of the neutrinos (provided  it has a mass of about 10 eV),
and the lightest supersymmetric particle, the neutralino, which is
expected to be stable. 
All these have to have a very weak interaction, so that they could not
be detected so far.

\subsection{\bf Gravitational microlensing}

The conclusion of {\em gravitational microlensing} of stars within  
the Large Magellanic Cloud (LMC) is
that MACHOs in the planetary mass range
$10^{-6}$ to 0.05 $M_\odot$ do not contribute a substantial fraction
of the Galactic dark halo.  In the two-year data of the LMC events
of the MACHO group 8 events could be detected which
are well in excess of the predicted background of approximately 1.1 events 
arising from known stellar populations. Hence,
MACHOs in a dark halo appear to be a natural explanation.

A statistical analysis of the {\em galactic halo} via microlensing 
\cite{Pa86,Su99} suggests that MACHOs account for a significant part 
($>$ 20\%)
of the total halo mass of our galaxy. Their most likely mass range
seems to be in the range 
0.3 -- 0.8 $M_\odot$, with an average mass of 0.5 $M_\odot$,
cf. \cite{J96,Su99}. 
If the bulge is more massive than the standard 
halo model assumes, the average MACHO mass \cite{J96} 
will be somewhat lower at 
$\sim 0.1$ $M_\odot$.

This can be viewed as an indication that 
MACHOs form an {\em distinct} large class of {\em old objects} 
that cannot be easily extrapolated from 
any familiar stellar population, such as brown or white dwarfs. 

However, there are some astrophysical difficulties with this
interpretation, mainly arising from
the estimated mass $\sim 0.5$ $M_\odot$ for the lenses.
These cannot be hydrogen-burning stars in the halo since
such objects are limited to less than 3\%\ of the halo mass by 
deep star counts \cite{gbf}.
Modifying the halo model to slow down 
the lens velocities can reduce the implied lens mass somewhat, but probably
not below the substellar limit 0.08 $M_\odot$.
Old white dwarfs have about the right mass and 
can evade the direct-detection constraints, but
it is difficult to form them with high efficiency, and there may be
problems with overproduction of metals and
overproduction of light at high redshifts from the luminous
stars which were the progenitors of the white dwarfs \cite{char-silk}. 
Primordial black holes are a viable possibility, though one has
to appeal to a coincidence to have them in a stellar mass range. 

Due to these difficulties of getting MACHOs in the inferred mass range 
without violating other constraints,
there have been a number of suggestions for explaining the
LMC events without recourse to a dark population:  
most of these suggestions construct some non-standard distribution of 
`ordinary' stars along the LMC line of sight.
However, these proposals appear somewhat contrived \cite{Su99}, but
can be tested observationally in the near future.

\subsection{Boson stars or axion stars as alternatives?}
It has been recently suggested \cite{J97} 
that MACHOs could rather be {\em primordial black holes} formed during
the early
QCD epoch in the inflationary scenario. For cosmological dark matter, 
bound states of gravitational waves, so-called 
`gravitational geons' built from spin--2 bosons, are also 
considered recently \cite{HMWW94}.

Since the standard model of elementary particles as well as their {\em
superstring} extensions involve also Higgs type scalar 
fields, we  will analyze here the alternative possibility 
that primordial {\em boson stars} account for this
non-baryonic part of dark matter \cite{sch97,sch98}. Boson stars are
descendants of the so-called {\em geons} of Wheeler \cite{W55,W61,W98},
except that they are built from  scalar particles (spin--0) 
instead of electromagnetic fields, i.e.~spin--1 bosons.  
If scalar fields exist in nature, such localized soliton-type configurations 
kept together by their self-generated gravitational field 
can form within Einstein's general relativity.  

In the case of {\em complex} scalar fields, an
{\em absolutely stable}  branch of such non-topological solitons 
with conserved particle number exists. In
the spherically symmetric case, we have shown via catastrophe theory
\cite{kus,SKM92} that these {\em boson stars} have a {\em stable
branch} with a wide range of masses and radii.

Kaup's  first 
investigation \cite{K68}
of the spherically symmetric boson star (BS)  was based on massive
scalar particles. Lateron, a nonlinear $U(|\Phi|^2)$ potential was introduced
by Mielke and Scherzer \cite{MS81}, where also solutions with nodes,
i.e.~``principal quantum number'' $n>1$, were found.
 In building macroscopic boson stars,  a nonlinear Higgs type  
self-interaction  
potential $U(|\Phi|^2)$ was later  considered \cite{colpi}
as an additional  repulsive interaction. 
Thereby the Kaup limit  for boson stars
can  even {\em exceed}  the limiting mass of 3.23 $M_\odot$
for  neutron stars \cite{RR74}. 

Three surveys \cite{jetzer,lee,Li92}
summarize the present status of the non-rotating case, a more
recent survey including the {\em rotating} BS  can be found in
\cite{mielkeMG8}.

Recently, we construct \cite{SM96,MS96} for the first time the 
corresponding  localized {\em differentially rotating} configurations 
via numerical integration of the coupled 
Einstein--Klein--Gordon equations. Due to {\em gravito--magnetic} effect,
the ratio of conserved angular momentum and particle number turns out 
to be an integer $a$, the {\em azimuthal quantum number} of our 
soliton--type stars. The resulting axisymmetric metric, the energy density 
and the Tolman mass are {\em completely regular}. 

\subsection{Are fundamental scalar fields part of nature?}
The physical nature  
of the  spin--0--particles out of which the boson star (BS) 
consists is still an open issue.
Until now, no fundamental elementary scalar particle has been found in 
accelerator experiments
which could serve as the main constituent of the boson star.
 In the electroweak theory of Glashow, Weinberg, and Salam, 
a {\em Higgs boson} dublett $(\Phi^+, \Phi^0)$ and its anti-dublett
$(\Phi^-, \bar \Phi^0)$ are necessary ingredients in order
to generate masses for the $W^{\pm }$ and $Z^0$ gauge vector bosons 
\cite{[5]}. After symmetry breaking, only one scalar particle, the
Higgs particle 
$h:=(\Phi^0 + \bar \Phi^0)/\sqrt{2}$, remains free and occurs as a state
in a constant scalar field background.  Nowadays,
calculations of the two--loop electroweak effects enhanced by powers 
of the mass of  the rather heavy top quark \cite{abe} 
has lead to an {\em indirect} determination of the Higgs mass, cf. 
\cite{G98}.
For $M_{\rm t}=173.8 \pm 5$ GeV/$c^2$, one 
finds   $m_{\rm h}= 104^{+93}_{-49}$ GeV/$c^2$. So far, the experimental 
constraints are weak; even for the unrealistic case of a Higgs mass   
up to  1000 GeV/$c^2$, the discrepancies for, e.g., the mass of the
$W$ boson would be small.
Above 1.2 TeV/$c^2$, however, the self--interaction $U(\Phi)$ of the 
Higgs field is so large that the perturbative approach of the 
standard model becomes unreliable. Therefore a conformal extension of 
the standard model with gravity included could be 
necessary, see \cite{hehl}. Fermilab's upgraded tevatron \cite{HZ99} 
has a mass 
reach of $135< m_{\rm h} < 186$ GeV/c$^2$, while the
high--energy experiments at the LHC at Cern will ultimately reveal 
if these Higgs particles really exist in nature. 

As free  particles, the Higgs boson
is unstable with respect to the decays $h \rightarrow W^+ + W^-$ and 
$h \rightarrow Z^0+ Z^0$. In an hypothetical compact
object like the BS, these decay channels are expected to be in 
equilibrium with the inverse process $Z^0+ Z^0 \rightarrow h$,
for instance.  Presumably, this is in full analogy with the neutron star
\cite{weigel} or quark star \cite{[BB94],weber,glenn},
where one finds an equilibrium of $\beta $-- and inverse $\beta $--decay 
of the neutrons or quarks and thus stability of the macroscopic star with 
respect to radioactive decay. Such Higgs sector nontopological
solitons \cite{BT95}
may also be candidates for cold dark matter.
Nishimura and Yamaguchi \cite{[6]}  constructed a
neutron star using an equation of state of an isotropic fluid built 
from Higgs bosons.

Nowadays there are many attempts of unifying the 
standard model with gravity on the quantum level, like 
string theory \cite{DP94}. Commonly, the four--dimensional 
{\em effective} models 
 make the  prediction \cite{Sh91} that 
the tensor field $g_{\mu\nu}$  of gravity is accompanied by one 
or several scalar fields. 

In string effective supergravities \cite{FKZ94},
the mass of the {\em dilaton} $\varphi$ can be related  to the 
supersymmetry breaking scale
$m_{\rm SUSY}$ by $m_\varphi 
\simeq 10^{-3}(m_{\rm SUSY}/$ TeV/c$^2)^2$ eV/c$^2$ with interesting astrophysical 
implications \cite{Da99}, but this is not the only 
possibility.

The other scalar field of the effective string Lagrangian is 
the `universal' {\em invisible axion} $\sigma$, a pseudo--scalar  
potential for the Kalb--Ramond three 
form $H:=e^\varphi\, ^*d\sigma$. 
(There are some speculations \cite{DT95,So98} to identify it with the
the axial part of a possible torsion of spacetime). 
{}From the Hubble scale 
$H_{\rm eq}\sim 10^{-27}$ eV/c$^2$ of matter--radiation equilibrium and 
 the temperature $T_{\rm m } \sim 100$ MeV of mass generation at 
the epoch of chiral symmetry breaking, one can derive  \cite{Ga99,GV99} 
the condition  $m_{\sigma} > (T_{\rm m}/{\rm eV})^2\,H_{\rm eq}$.
This allows a very light axion mass 
$m_{\sigma} = 7.4\times (10^{7}{\rm  GeV}/f_{\sigma})$ eV/c$^2$
$>10^{-11}$ eV/c$^2$ with
decay constant $f_{\sigma}$ close to the inverse Planck time, thus a prime
candidate for dark matter. (This should not be confused with  the 
Goldstone boson $a$ of the 
Peccei--Quinn symmetry \cite{Pe89} of standard QCD, for which
a recent experiment \cite{HKSB98} has
excluded the range of $m_{\rm a}\sim 10^{-6}$ eV/c$^2$. From cooling
neutron stars, there
can be inferred \cite{UITQN} an upper limit of $m_{\rm a}< 0.06$ --
0.3 eV/c$^2$, depending on the equation of state of the nucleon fluid.

\section{Boson stars}

In a 1968 perspective paper, Kaup \cite{K68} has studied for the first time  
the full generally relativistic coupling of linear Klein--Gordon fields 
to gravity in a {\em localized configuration}. This 
`Klein--Gordon geon' is nowadays  christened
{\em mini--boson star} and can be regarded  as a 
{\em macroscopic quantum state}.
It was already realized that  
{\em no} Schwarzschild type 
event horizon occurs in such  numerical solutions. Moreover, the problem of 
the stability of 
the resulting scalar `geons' with respect to radial perturbations is treated. 
It is shown that such objects are, below a well-defined critical mass,  
resistant to gravitational collapse. 
These considerations are on a semiclassical level, since the
Klein--Gordon field is treated as a classical field. 

The Lagrangian density of gravitationally coupled complex scalar 
field $\Phi$ reads
\be
{\cal L}_{\rm BS} = \frac{\sqrt{\mid g\mid }}{2\kappa } 
\{ R+ \kappa [ g^{\mu \nu} (\partial_\mu \Phi^*) 
(\partial_\nu \Phi )- U(\mid \Phi \mid^2)]\} \, ,  
             \label{lagrange}
\ee
where $\kappa= 8\pi G$ is the gravitational constant.
Using the principle of variation, one finds the coupled
Einstein--Klein--Gordon equations
\ba
 G_{\mu \nu }:=  R_{\mu \nu } - \frac{1}{2} g_{\mu \nu } R
             & = & -\kappa T_{\mu \nu } (\Phi ) \; , \label{phi152} \\
      \left (\Box + \frac {dU}{d\mid \Phi \mid^2} \right ) \Phi
             & = & 0 \; , \label{phi153}
\ea
where
\be
T_{\mu \nu }(\Phi )
 = {1\over2} [ (\partial_\mu \Phi^\ast )(\partial_\nu \Phi )
  + (\partial_\mu \Phi )(\partial_\nu \Phi^\ast ) ]
  - g_{\mu \nu } {\cal L} (\Phi )/\sqrt {\mid g\mid }
\ee
is the {\em stress--ener\-gy ten\-sor} and
$
\Box := \left (1/\sqrt {\mid g\mid }\right )\,
 \partial_\mu  \left (\sqrt{\mid g\mid } g^{\mu \nu }
 \partial_\nu \right )
$
the generally covariant d'Alembertian.

\subsection{Spherically symmetric solutions}
The stationarity ansatz
\be
\Phi (r,t)=P(r) e^{-i\omega t} 
\ee
 describes a spherically symmetric bound 
state of the scalar field with frequency $\omega $.

In the case of spherical symmetry and isotropic cordinates, 
the line-element reads
\be
ds^2 = e^{\nu (r)} dt^2 - e^{\lambda (r)} \biggl [ dr^2 + r^2 \Bigl (
 d\theta^2 + \sin^2\theta d\varphi^2 \Bigr ) \biggr ] \; ,
\ee
in which the functions $\nu =\nu (r)$ and 
$\lambda =\lambda (r)$ depend on the radial
coordinate $r:= \sqrt{x^2 +y^2+z^2}$.

In a nut--shell, a boson star is a stationary solution of a 
(non-linear) Klein--Gordon equation  in its own 
gravitational field; cf.~\cite{Mi78,Mi80a}.
As in 
the case of a prescribed Schwarzschild background \cite{DM79}, 
the spacetime  
curvature affects the resulting {\em radial  Schr\"o\-din\-ger equation}
\be \left[\partial_{r*^2} -V_{\rm eff}(r) +\omega^2- m^2\right]P=0
\ee for the
radial function $P(r)$ 
essentially via an {\em effective} gravitational 
potential $V_{\rm eff}(r)$, when written in terms of 
 the tortoise coordinate $dr^*:= e^{(\lambda- \nu)/2}\, dr$  .
Then it can be easily  realized that localized solutions 
fall off asymptotically as 
$P(r) \sim (1/r)\exp\left(-\sqrt{m^2 -\omega^2}\, r\right)$ in a 
Schwarzschild-type asymptotic background.

The energy--momentum tensor becomes diagonal,
i.e.~$T_\mu{} ^\nu(\Phi) = {\rm diag} \; (\rho , -p_r,$
$-p_\bot, -p_\bot )$ with
\ba
\rho &=& \frac{1}{2} (\omega^2 P^2 e^{-\nu} +P'^2 e^{-\lambda} +U )\; ,
\nonumber  \\
p_r &=&  \rho -  U \; , \;  \nonumber  \\ 
p_\bot &=&  p_r -  P'^2 e^{-\lambda } \; .   
 \ea
This  form  is familiar from  fluids,
except that the radial and tangential pressure generated by the
scalar field are in general different, i.e.~$p_r \neq p_\bot $, 
due to the different sign of  
$(P')^2$ in these expressions. 

In general, the resulting system of three coupled nonlinear equations for the
radial parts of the scalar and the (strong) gravitational tensor field has
to be solved numerically. (Exact solution of massless scalar fields 
\cite{BMHH87} or of 
the coupled Maxwell--Einstein--Klein--Gordon equation \cite{D63} 
tend to be plagued with 
singularities.)
 
 In order to specify the starting values for the 
ensuing numerical analysis, asymptotic solutions at the
origin and at spatial infinity are instrumental.  The resulting 
configuration turns out to be 
{\em completely regular} and does {\em not} exhibit an 
apparent event horizon, cf. \cite{MS96}.

The stress--energy tensor of  a 
BS, unlike a classical fluid,  is in general {\em anisotropic} \cite{K68}. 
In contrast to neutron stars \cite{HT65,[Zel]}, where the ideal fluid
approximation demands an
isotropic symmetry for the pressure, for spherically symmetric 
boson stars there are different stresses $p_r$ and $p_\bot$  in radial or 
tangential directions, respectively. The
{\em fractional anisotropy}\ $a_f :=(p_r - p_\bot )/p_r =
P'^2 e^{-\lambda}/(\rho -U)$  depends essentially 
on the self-interaction; cf.~Ref.~\cite{RB69,Ge88}.
So, the perfect fluid approximation is inadequate for boson stars. 

There exists a decisive difference between self--gravitating 
objects made of fer\-mions or bosons:
For a many fermion system the Pauli exclusion principle 
forces the typical fermion 
into a state with very high quantum number, whereas many bosons can 
coexist all in the same ground state (Bose--Einstein condensation). 
This also reflects itself 
in the critical {\em particle number} $N:=\int_0^\infty j^0 dr$ 
of stable configurations:
\begin{itemize}
\item{}
$N_{\rm crit} \simeq (M_{\rm Pl}/m)^3$ for fermions 
\item{}
$N_{\rm crit} \simeq (M_{\rm Pl}/m)^2$ for massive bosons without 
self--interaction. 
\end{itemize}  

Cold mixed boson--fermion stars have been studied by 
Henrique et al.~\cite{HLM89} and 
Jetzer \cite{jetzer}.

\subsection{Critical masses of boson stars}
Since boson stars are {\em macroscopic quantum states}, below a
certain {\em critical} mass $M_{\rm crit}$ they are 
prevented from complete
gravitational collapse by the Heisenberg uncertainty principle 
$\Delta r \Delta p \sim \pi \hbar$, cf. Ref.\cite{Lieb}.
This provides us also with crude mass estimates: For a boson to be confined 
within the star of {\em effective}  
radius $R_{\rm eff}:=(1/N)\int_0^\infty j^0 rdr$, the Compton
wavelength of the collective boson has to satisfy
$\lambda_\Phi= (2\pi\hbar/mc) \leq 2R_{\rm eff}$. On the other hand,
the star's radius    
should be of the order of the last stable Kepler orbit 
$3R_{\rm S}$ around a black hole of Schwarzschild radius  
$R_{\rm S}:= 2GM/c^2$ in order to avoid an instability 
against complete gravitational collapse.

For a {\em mini-boson star}, i.e. a massive boson model 
with merely the mass term 
$U(\vert\Phi\vert) =m^2\vert \Phi\vert^2$ as self-interaction, of an effective  
radius  $R_{\rm eff} \cong (\pi/2)^2 R_{\rm S}$ close to 
the last stable Kepler orbit of a black hole,  
one obtains the estimate
\be 
M_{\rm crit} \cong (2/\pi)M_{\rm Pl}^2/m \geq  0.633\, 
M_{\rm Pl}^2/m =M_{\rm Kaup}\, ,  \label{Kauplim}
\ee
cf.~Ref.~\cite{jetzer,mielkeMG8}. This provides us with 
a rather good upper bound on the so-called {\em Kaup limit}.
The correct value in the second expression 
was found  numerically \cite{K68} as a limit of  the 
maximal or {\em critical} mass of a {\em stable} mini--BS.
Here $M_{\rm Pl}:=\sqrt{\hbar c /G}$ is the Planck mass
and $m$ the mass of a bosonic particle. 

For the likely  mass $m_{\rm h}=100$ GeV/c$^2$ of the Higgs particle, e.g.,
one can estimate  
the total mass of this mini-boson star to be $M\simeq 10^{10}$ kg and its
radius $R_{\rm eff}\simeq 10^{-18}$ m yielding a
density 10$^{45}$ times that of a neutron star.

A boson star is an extremely dense object, since 
non-interacting scalar matter is very ``soft".  However, these   
properties are changed considerably by considering a
{\em repulsive} self-interaction
\be
U(\vert\Phi\vert) = m^2\vert \Phi\vert^2\left(1 + \frac{1}{8} 
\Lambda(\vert\Phi\vert^2)\right) =
m_{\rm ren}^2\vert \Phi_{\rm ren}\vert^2 \,.
\ee
where $\Lambda(\vert\Phi\vert^2)$ denotes an arbitrary 
nonlinear self-interaction. The 
choice $\Lambda(\vert\Phi\vert^2)= (4\lambda/m^2)\vert\Phi\vert^2$ 
would lead us back to the
quartic self-interaction
of Colpi et al. \cite{colpi}. If we
adopt the value $ \vert \Phi_0\vert\simeq M_{\rm Pl}/\sqrt{16\pi}$
inside the boson star, one finds for the energy density
\be
\rho \simeq m^2 \vert \Phi_0\vert^2\left(1 + \Lambda/8\right) \, , 
\ee
 where $\Lambda:= \Lambda(M_{\rm Pl}^2/16\pi)$ 
is a dimensionless 
coupling constant such that  we would recover 
$\Lambda:= (\lambda M_{\rm Pl}^2 /4\pi m^2)$ for the quartic interaction.
The self-interaction becomes dominating for $\Lambda \geq 8$ or
$\lambda \geq 32\pi (m/M_{\rm Pl})^2$. Thus, even   a rather
tiny coupling constant $\lambda$ could have drastic effects on a BS.
 
Formally, this corresponds to a star formed from non-interacting bosons
$\Phi_{\rm ren}= \Phi(1 + \Lambda/8)$
with a {\em lower} renormalized mass 
$m\rightarrow m_{\rm ren}:=m/\sqrt{1 +  \Lambda/8} $ but 
{\em larger} Compton wave length $\lambda_{\Phi({\rm ren})}$ and,
consequently, a larger  
radius $R_{\rm eff}$.  (A reverse rescaling of the mass, as was presumed 
in a recent preprint \cite{HKL99}, 
leads to a smaller Compton wave length and other inconsistencies.) 
Consequently, we can apply again (\ref{Kauplim}) for the Kaup limit
and find that
the maximal mass of a  {\em stable} BS scales approximately as 
\be 
M_{\rm crit}\simeq (2/\pi)M_{\rm Pl}^2/m_{\rm ren} ={2\over\pi}\sqrt{1
+  \Lambda/8}
{M_{\rm Pl}^2 \over m} \; \rightarrow \;
 {1\over{\pi\sqrt{2}}}\sqrt{\Lambda} {M_{\rm Pl}^2 \over m} \quad
{\rm for}\;  \Lambda \rightarrow \infty
\, . \label{Colpimass}
\ee
For the quartic self-interaction, this accounts
rather well for  the numerical results of Colpi et al. 
\cite{colpi}. Our formula not only reproduces their 
asymptotic mass  formula (11) for $\Lambda\rightarrow\infty$, 
but, by construction,  interpolates as well with the
Kaup limit (\ref{Kauplim}) for $\Lambda=0$.

$$\vbox{\offinterlineskip
\hrule
\halign{&\vrule#&\strut\quad\hfil#\quad\hfil\cr
height2pt&\omit&&\omit&&\omit&\cr
& {\bf Compact} && {\bf Critical mass} && {\bf Particle Number } &\cr
&{\bf Object}  && $M_{\rm crit}$ && $N_{\rm crit}$ &\cr
height2pt&\omit&&\omit&&\omit&\cr
\noalign{\hrule}
height2pt&\omit&&\omit&&\omit&\cr
& Fermion Star:  &&  $M_{\rm Ch}:=M_{\rm Pl}^3/m^2$  &&   $\sim(M_{\rm
Pl}/m)^3$ &\cr
& Mini--BS:  &&  $M_{\rm Kaup}=0.633\, M_{\rm Pl}^2/m $  &&   $0.653 \,(M_{\rm Pl}/m)^2$ &\cr 
& Boson Star:  &&  $(1/\sqrt{2\pi})^3\sqrt{\lambda} M_{\rm Ch}$  
&&   $\sim (M_{\rm Pl}/m)^3$ &\cr 
& Soliton Star:\cite{L87,lee}  
&&  $10^{-2} (M_{\rm Pl}^4/ m \Phi_0^2)$  &&   
$2\times 10^{-3} (M_{\rm Pl}^5/ m^2 \Phi_0^3)$ &\cr 
height2pt&\omit&&\omit&&\omit&\cr}
\hrule}$$

The {\em Chandrasekhar limit} is  
$M_{\rm Ch}:=M_{\rm Pl}^3/m^2\simeq 1.5({\rm GeV/mc}^2)^2$ $M_\odot$,
where $M_\odot$ denotes the solar mass. In
astrophysical terms, the maximal BS
mass is $M_{\rm crit} \cong 0.06\sqrt{\lambda}\, M^3_{\rm Pl}/m^2$ $=0.1
\sqrt{\lambda}$ ({\rm GeV/mc}$^2)^2$ $M_\odot$ which for $\lambda=1$ and 
proton mass $m \simeq 1$ GeV/c$^2$ is in the 
interesting mass range $\sim 0.1$ $M_\odot$ of MACHO's.

In a scale-invariant theory built from 
nonlinearly coupled dilatons $\varphi$, there arise a conserved 
{\em dilaton charge} via Noether's theorem 
from Weyl rescaling and thus will ensure the stability of the 
configuration. For a {\em dilaton star} with  quadratic
self-interaction \cite{TX92}, the same formula (\ref{Colpimass})
applies, but  the  coupling constant 
$\Lambda:= (\lambda M /4\pi \omega)^2$ will be $\omega$ dependent. For
a very light dilaton $\varphi$  of mass
$m_{\rm dil}=10^{-11}$ eV/c$^2$, resembling  a misaligned  
`universal' axion at its lower mass bound, Gradwohl and K\"albermann \cite{GK89} found 
\be
M_{\rm crit} = 7\sqrt{\overline\lambda}\, M_\odot\, , \qquad   
R_{\rm crit} = 40\sqrt{\overline\lambda}\; {\rm km}\,,
\ee
where $\overline\lambda:= \lambda(M/\omega)^2$ is the rescaled
coupling constant of the $\varphi^4$ interaction.

To repeat, in building macroscopic boson  a 
Higgs--type self-interaction $U(\vert\Phi\vert)$ is crucial for 
accommodating a repulsive force besides gravity. This
repulsion between the constituents is instrumental to
blow up the boson star so that much more particles will have
room in the confined region. Thus the maximal mass of a BS 
can reach or even extend the limiting mass of 3.23 $M_\odot$
for  neutron stars \cite{RR74,fried,cook} with {\em realistic} equations of
state $p=p(\rho)$ for which
the (phase) velocity of sound is $v_s=\sqrt{dp/d\rho}\leq c$. However,
this fact depends on the strength of the self--interaction.

Therefore, if scalar fields would exist in nature, such compact objects  
could even question the observational AGN black hole paradigm in  
astrophysics.

\section{Excited boson stars}
\subsection{Gravitational atoms as boson stars}
Ruffini and
Bonazzola \cite{RB69,[4]} used the  formalism of  second quantization 
for the complex Klein--Gordon field and noticed  
an important feature: If all
scalar particles are within the {\em same ground state}
$|\Phi \rangle =(N,n,l,a)=(N,0,0,0)$, which 
is possible
because of Bose--Einstein statistics, then the {\em semi--classical}
Klein--Gordon equation of Kaup can be  {\em recovered} in the 
Hartree--Fock approximation for the second quantized two--body problem. 
  In contrast to the Newtonian 
approximation, the full relativistic 
treatment avoids an unlimited increase of the 
particle number $N$ and negative energies, but induces 
critical masses and particle numbers with a global maximum.

Due to this Hartree--Fock approximation  and while also neglecting 
 effects of the quantized gravitational field,
the same coupled 
Einstein--Klein--Gordon equations (\ref{phi152},\ref{phi153}) apply.  
 Therefore, a  boson star is also 
 called a {\em gravitational atom} \cite{FG89}. Since a 
 free Klein--Gordon equation for a complex scalar field is a 
 {\em relativistic generalization} of the 
 {\em Schr\"odinger equation}, we consider for the {\it ground state} 
 a generalization of the wave function
 \ba
 \vert N, n, l,a\rangle: \qquad \Phi&=& R_a^n(r)\, 
 Y^{a}_l(\theta, \, \varphi)e^{-i (E_n/\hbar)t} \nonumber\\
 &=&{1\over\sqrt{ 4\pi}} R_a^n(r)
 P^{a}_l(\cos\theta)\, e^{ia\varphi}\, 
e^{-i (E_n/\hbar)t} \label{Hatom}
\ea 
of the hydrogen atom. Here $R_a^n(r)$ is the radial distribution,
$Y^{a}_l(\theta, \, \varphi)$ the spherical harmonics,  
$P^{a}_l(\cos\theta)$ are the 
normalized Legendre polynomials, and $|a|\leq l$ are  the quantum numbers
of {\em azimuthal}  and {\em angular} momentum.
 
Due to their inherent `gravitational confinement' gravitational atoms 
represent {\em coherent} quantum states, which
nevertheless can have macroscopic size and large masses. 
The gravitational field is self-generated 
via the energy--momentum tensor, but remains 
completely classical, whereas the complex
scalar fields are treated to some extent as Schr\"odinger 
wave functions, which in quantum field theory are 
referred to as semi-classical.

Moreover, Feinblum and McKinley \cite{FM68} found eigensolutions with
nodes corresponding to the {\em principal 
quantum number} $n$ of the H--atom.  
Motivated by Heisenberg's non-linear spinor equation \cite{Mi81}
additional self--inter\-acting terms   describing the interaction 
between the bosonic particles in a ``geon'' type configuration were first 
considered by  Mielke and Scherzer \cite{MS81}, where also solutions
with nodes, i.e.~``principal quantum number'' $n>1$ and non-vanishing
angular momentum  $l \neq 0$ 
for a t'Hooft type monopole \cite{tH74} ansatz $\Phi^{I=a} \sim R(r)\, P^{|a|}_l(\cos\theta)$ 
were found. These highly interesting
instances of a possible fine structure in the energy levels of 
gravitational atoms poses the question if
{\em quantum geons} \cite{W61,W98} are 
capable of internal excitations?
Recently, without reference to these earlier works, such 
{\em ``exited boson stars"} were recovered \cite{[B2],BSS98} 
and their stability properties corroborated in some numerical
details. Moreover, Rosen \cite{rosen} reviewed his old idea of an elementary 
particle built
out of scalar fields  within the framework of the Klein--Gordon geon  (or 
the mini--boson star, as they are christened today). 

Several  surveys \cite{jetzer,lee,strau,mielkeMG8}
summarize the present status of the non-rotating case.

\subsection{Rotating boson stars} 
In the framework of 
Newtonian theory, boson stars with axisymmetry have been  
constructed by several groups.
Static axisymmetric boson stars, in the Newtonian limit \cite{SB96}
and in GR \cite{YE97}, show that one can distinguish
two classes of boson stars by their parity transformation at 
the equator. In both approaches only the negative parity solutions
reveal axisymmetry, while those with positive parity merely  
converged  to
solutions with spherical symmetry.
The metric potentials and the components of the energy-momentum tensor
are equatorially symmetric despite of the antisymmetry of the scalar field.
In the Newtonian description, Silveira and de Sousa \cite{SS95} 
followed the approach of Ref.~\cite{FG89} and constructed
solutions which have no equatorial symmetry at all. Hence, in GR, 
we have to separate solutions with and without equatorial symmetry.

Kobayashi et al. \cite{koba} tried to find slowly rotating states
({\em near} the spherically symmetric ones) of general relativistic
boson stars, but they failed.
The reason for that is a quantization of 
the {\em total angular momentum} \cite{wini}
\be
J = \int T_3{}^0\,\sqrt{|g|} d^3 x = a N 
\qquad\qquad a= 0, \pm 1, \pm 2, \cdots\label{angular} \; 
\ee
of boson stars which  is proportional to the particle number $N$ and 
vanishes if $a=0$. This relation between 
angular momentum and particle number was first derived by 
Mielke and Schunck \cite{MS96}. 

In recent papers \cite{SM96,miesch}, we proved numerically that 
{\em rapidly rotating} boson stars with $a\neq 0$  
exist in general relativity. 
Because of the finite velocity of light and the infinite range of the
scalar matter within the boson star, our
{\em localized} configuration can {\em rotate only differentially}, but not
uniformly. This new axisymmetric solution of the 
coupled Einstein--Klein--Gordon equations represent the 
{\em field-theoretical pendant} of 
rotating neutron stars which have been studied numerically 
for various equations of state and different approximation schemes 
\cite{fried,cook,[Er93],Bo93} as well as for differentially 
rotating superfluids \cite{LSC98} as a model for (millisecond) pulsars.

On the basis of Ref. \cite{koba}, it has erroneously been  claimed
\cite{YE97} that ``rapidly rotating boson stars cannot exist".
However, more recently \cite{YE97b}
the same Japanese group (as well as \cite{R97}) followed exactly  
our Ansatz  and could verify 
all our earlier results \cite{SM96,MS96}, albeit of 
some extension to stronger gravitational fields, due 
to better computational facilities. Due to the anisotropy of the
stress--energy tensor, our configuration is {\em differentially rotating},
see \cite{MS96} for more details.

Moreover,  the energy density of our rotating boson star is
concentrated in an effective {\em mass torus} \cite{miesch}. Thus this first 
{\em nonsingular} model of a {\em rotating body}  in GR
realizes to some extent the suggestion of Newman et al. \cite{NJ} to 
fill in the Kerr metric, in view of its ring singularity, with a 
{\em toroidal} rather than a spherical {\em source}.  
Toroidal structure occurs also in relativistic star systems 
with an accretion disk \cite{NEL92}.

Since rotating BS have a toroidal structure, there seems to exist the more speculative 
possibility of {\em knotted} vortex like excitations, cf. Ref. \cite{Mi77}. 
For an $O(3)$ Skyrme model, their existence has recently been demonstrated numerically 
\cite{FN97,BS98}.

\subsection{Formation of (primordial) boson stars}
The possible abundance of solitonic stars with astrophysical 
mass but microsco\-pic size could have  interesting implications 
for galaxy formation,  the
microwave back\-ground, and formation of protostars. 
The formation of non-gravitating non-topological solutions was 
already studied by Frieman et al.~\cite{FGGK}.)
In comparison with primordial black holes,
 it is therefore  an important question if boson 
stars can actually form from a primordial bosonic ``cloud" \cite{T91}. 
 
Collisionless star 
systems are known to settle to a centrally denser system by sending some  
of their members  to larger radius. Likewise, a bosonic cloud will settle to a 
unique boson star by ejecting part of the scalar matter. Since there is no 
viscous term in the KG equation (\ref{phi153}), the `radiation  of 
the scalar field is the
only  dissipationless relaxation process  called 
{\em gravitational cooling}. Seidel and Suen \cite{SS90,SS94} demonstrated this 
 numerically 
by starting with a spherically symmetric configuration with 
$M_{\rm initial} \geq M_{\rm Kaup}$, i.e.~which is more massive then 
the Kaup limit. Actually such oscillating and pulsating branches have been 
predicted earlier in the stability analysis of Kusmartsev, Mielke, and 
Schunck \cite{KMS91,SKM92,KS92} by using 
{\em catastrophe theory}. Oscillating soliton stars were constructed by using
real scalar fields which are periodic in time \cite{SeSu91}.
Without spherical symmetry, i.e.~for 
$\Phi \sim R_a(r) Y_l{}^a(\theta\, ,\varphi) $, the emission of 
gravitational waves would also be necessary. 

 For the formation of {\em primordial} BSs, an important issue is the breaking of
unified gauge (super--)symmetry at high temperature  in order to 
 yield a scalar-antiscalar asymmetry $\epsilon_S = N_\Phi/N_\gamma$,
as in the case of baryon-antibaryon asymmetry, where 
$\epsilon_B = N_B/N_\gamma \sim 10^{-10}$.
Here, we recall the situation of collapsing homogeneous mini-BS
clouds in the early Universe, cf.~\cite{khlo85,Li92}.
Because the Jeans scale at decoupling
time is greater than the horizon scale, a bosonic mass of
$M_{{\rm  Pl}}^3/m^2$ immediately collapses and since this is a
factor $M_{{\rm  Pl}}/m$ higher than the maximal mass allowed within the
mini-BS model, only black holes can form.
For an asymmetry factor of order $\epsilon_S \sim m/M_{{\rm  Pl}}$, however,
the total mass remaining within the horizon is
$M_{{\rm  Pl}}^2/m$, hence, BSs could form, avoiding the final state of a
black hole.

For a real (pseudo-)scalar field, like the axion $a$
of the broken Peccei--Quinn symmetry \cite{Pe89} 
in QCD, the outcome is
quite different. 
The axion has the tendency to form compact objects 
(oscillatons) in a short time scale. Due to its intrinsic oscillations 
it would be unstable, contrary to a BS. Since the field disperses to 
infinity, finite non-singular self--gravitating solitonic objects
cannot be formed with a massless Klein--Gordon field \cite{Ch86,P87}.
In Ref. \cite{KT93} a different mechanism for forming {\em axion miniclusters} 
and starlike configurations was proposed. The self--coupling relaxion time 
\cite{T91} is compatible or larger as the 
age of the Universe. For fermionic soliton stars, there is a
 temperature dependence \cite{CV91} in the forming of cold configurations. 

\subsection{Gravitational waves} 
In the last stages of boson star  formation, one 
expects that first a highly excited configuration forms in which 
the  quantum numbers $n$, $l$ and $a$ of the gravitational atom, 
i.e.~the number $n-1$ of nodes, the 
angular momentum and the azimuthal angular dependence $e^{ia\varphi}$
are non-zero.  

In a simplified picture of BS formation, all initially high 
modes have eventually to decay 
into the ground state $n=l=a=0$  by a {\em combined emission} of scalar 
radiation and gravitational radiation.

In a Newtonian approximation of Ferrell and Gleiser \cite{FG89}, the energy released 
by scalar radiation from states with  zero quadrupole moment 
can be estimated by
\be E_{\rm rad} \sim (n-1) M_{\rm Pl}^2/m\, .
\ee
This is accompanied by a loss of boson particles with 
the rate $\Delta N\sim (n-1)  (M_{\rm Pl}/m)^2$.
For investigating the gravitational
radiation of macroscopic boson stars with large self-interaction, a 
reduction of the differential equations 
can be  taken into account \cite{R97}.

The lowest BS mode which has {\em quadrupole moment} and therefore 
can radiate {\em gravitational waves} is the $3d$ state 
with $n=3$ and $l=2$. For $\Delta J=2$ transitions,  it will decay into 
the $1s$ ground state with $n=1$ and $l=0$ while preserving the 
particle number $N$.   The radiated energy is 
quite large, i.e., $E_{\rm rad} =2.9 \times 10^{22}$ (GeV/mc$^2$) Ws.
Thus the final phase of the BS formation would terminate in an 
outburst of gravitational radiation despite the smallness of the object.

\subsection{Gravitational evolution}

There would occur an {\em evolution}
of boson stars if the external gravitational constant $\kappa$ 
changes its value
with time \cite{GJ93,T97,CS97,TLS97,torr2}.
This can be outlined within the theory of Jordan--Brans--Dicke
or a more general scalar tensor theory. The results show that
the mass of the boson star decreases due to a space-depending
gravitational constant, given through the Brans--Dicke scalar.
The mass of a boson star with constant central density is influenced
by a changing gravitational constant. Moreover, the possibility of a
gravitational memory of boson stars or a formation effect upon their
surrounding has been analyzed as well \cite{TLS97}.

\section{Are MACHOs axion stars?}
Direct observation of boson
stars seems to be impossible also in the far future. We propose here
several effects which could possibly give indirect evidence \cite{SL97}.
In the asymptotic region, the rotation velocity of
baryonic objects surrounding the boson star can reveal the star's mass.
Assuming that the scalar matter of the BS interacts mainly
gravitationally, we would have a `transparent' BS detecting a
gravitational redshift up to values of $z=0.68$ observable by
radiating matter moving
in the strong gravitational potential. For further investigations of
rotation curves, cf.~Ref.~\cite{Sin95,LK96}, and of boson stars as
gravitational lenses, cf.~\cite{dabr}.

Solutions with an infinite range can be found where the
mass increases linearly \cite{sch97,sch98}. In the context of 
the dark matter hypothesis, it may be speculated if such {\em boson halos}
as well as excited BS states \cite{LK96,Sin95} can be
used to fit the observed rotation curves for dwarf and spiral galaxies
\cite{sch97,sch98}. Boson halos have a finite radius if a positive
cosmological constant exists as most recent results from supernovae
reveal \cite{Perlmutter98,riess98}.

Moreover, BSs could be the solution for the MACHO problem, as we are 
going to analyze in more detail.

In  {\em effective 
string} theories, the dilaton
$\varphi$, another moduli field $\beta$, and the  `universal' 
invisible axion $\sigma$ are predicted \cite{Sh91}.
This  can be read off 
from the effective string Lagrangian
\be
{\cal L}_{\rm eff} = \sqrt{\mid g\mid }e^{-\varphi} 
\left[ R + g^{\mu \nu}\left(\partial_\mu \varphi\partial_\nu \varphi -
6 \partial_\mu \beta\partial_\nu \beta -
\frac{1}{2} e^{2\varphi}\partial_\mu \sigma\partial_\nu \sigma\right) 
-2 \Lambda\right] \, .  \label{efflag}
\ee 
This corresponds to Eq. (11) with $\eta=2$ of Ref. \cite{DOT95} and            
 allows to 
combine \cite{Se93} the axion and the 
dilaton into a  {\em single complex} scalar field 
$\Phi:= \sigma +ie^{-\varphi}$, the {\em axidilaton}. In the
conformally related  Einstein frame 
$g_{\mu\nu} \rightarrow \widetilde g_{\mu\nu} :=e^{-\varphi}\,g_{\mu\nu}$  
and for constant modulus $\beta$, 
our results on BSs can easily by transferred 
to this {\em axidilaton} content of strings. 
\subsection{Mass range of axion stars}
As {\em macroscopic quantum states},  BSs are quite generally 
prevented from complete
gravitational collapse below a critical total mass $M_{\rm crit}$ which, 
 typically, depends inversely on the particle mass, see
Eq. (\ref{Kauplim}). 

\begin{figure}[bht]
\centering  
\leavevmode\epsfysize=8 cm \epsfbox{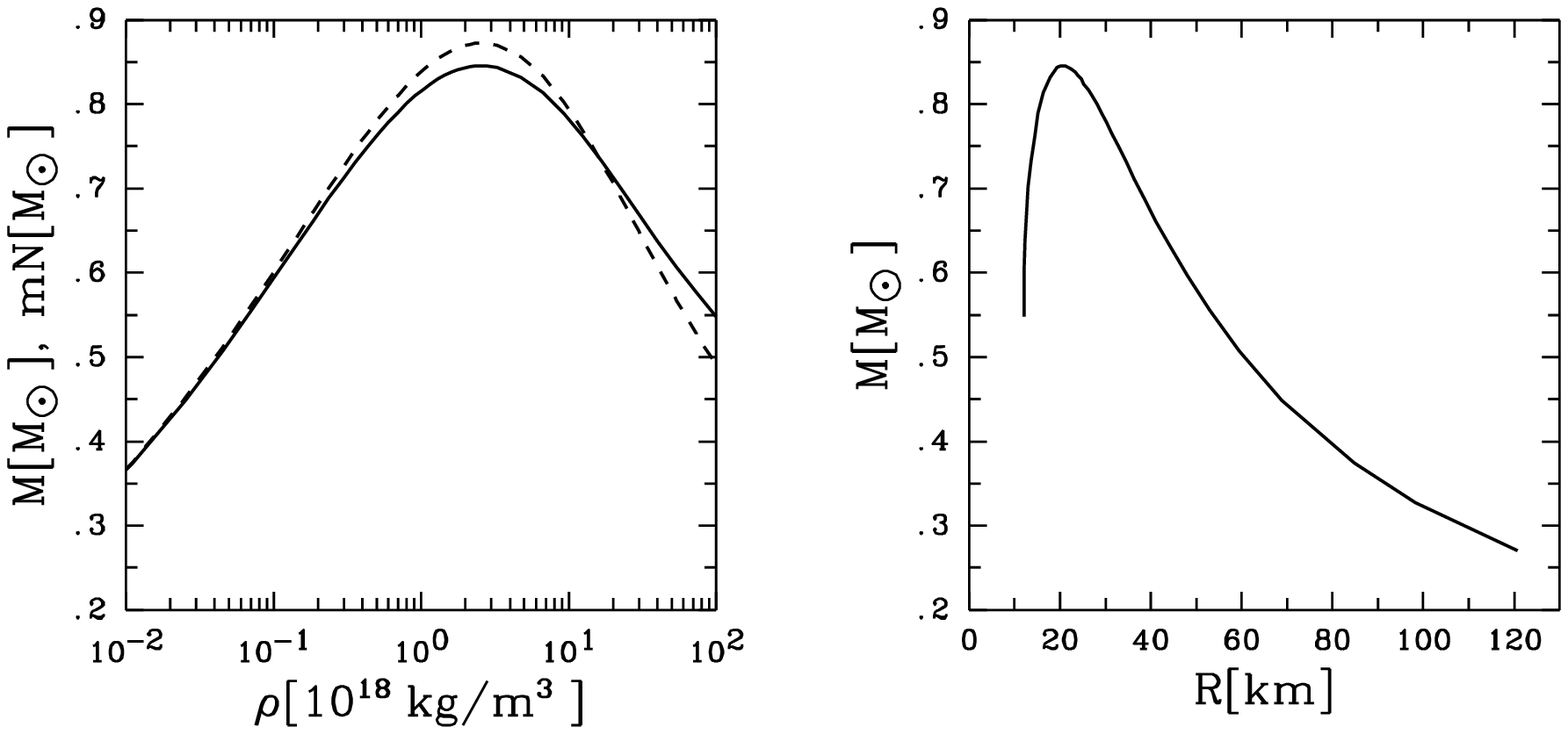}\\
\caption[fig1]{Left: Mass $M$ and particle number $N$ (or
rest mass $mN$ at infinity) of a BS depending on the central density
$\rho$. Right: Mass--radius dependence of an axionic BS.\label{fig1}} 
\end{figure}

The numerical results are shown in Fig.\ref{fig1}.
The left figure exhibits the dependence of  
the mass $M$ and the particle number $N$ (rest mass)
on the central density $\rho_0$. 
{\em Stable axionic} BSs exist at 
central densities  lower than the maximum mass.
The critical values are: $M_{\rm crit}=0.846$ $M_\odot$,
$mN_{\rm crit}=0.873$ $M_\odot$ and $\rho_{\rm c}=9.1\times \rho_{\rm
nucl}$, where $\rho_{\rm nucl}=2.8\times 10^{17}$ kg/m$^3$ is the average
density of nuclei. Since
non-interacting bosons are very ``soft", BSs are extremely dense objects
with a critical density higher than for neutron or strange stars 
\cite{glenn}. The figure on the right hand side gives the mass
depending on the radius (measured in km). 
For the mass--radius diagram, we have chosen as radius
99.9\%\ of the total mass. This ensures that the exponentially 
decreasing `atmosphere' of the
BS has almost no influence on the
asymptotic Schwarzschild spacetime.

The stable BSs or {\em axion stars} (ASs) 
have radii larger than the minimum at 20.5 km 
and a mass of 0.846 $M_\odot$.
In order to derive these values, we have assumed that the 
mass of the scalar field is $10^{-10}$ eV
close to the lower bound of {\em axions}, leading
to an asymmetry factor of $\epsilon_S \sim 10^{-38}$. and that no
self-interaction exists. 
We stress that the total mass of these relativistic  ASs is
just in the observed range of 0.3 to 0.8 $M_\odot$ 
for MACHOs. One could also turn this argument around: By identfying 
the MACHOs with known gravitational mass 
of about 0.5 $M_\odot$ with ASs, we are essentially `weighing", via 
$M_{\rm Kaup}/N_{\rm crit} \cong m$, 
 the axion mass to  $m_\sigma\sim 10^{-10}$ eV/c$^2$. It is gratifying to note that 
 such a low value 
 is perfectly compatible with the constraints on the 
 mass range  of the Kalb--Ramond axion
seeding the large-scale CMB anisotropy, cf. the recent results of 
 Gasperini and  Veneziano \cite{GV99,Ga99} within low-energy string cosmology.

For the other option of  
 dilaton $\varphi$ being
{\em stabilized} \cite{Di97} through the axion,
 a much smaller dilaton mass of $m_\varphi \sim 10^{-6}\; m_{\rm a}$ 
could  be generated non-perturbatively, such that 
the dilaton behaves very similar to 
misalignment produced Peccei--Quinn axion $a$.
Our conclusion also with respect to 
 the mass range of an {\em axidilaton star} 
will not changed much, if we  
use the full Brans-Dicke type interaction
\cite{GK89} for the combined axidilatons. 
  
Thus, for cosmologically relevant (invisible) axions as cosmological 
dark matter also an AS \cite{T91,SL97,KT93} with 
a rather large mass of would be possible and stable.

Therefore, if such--string inspired scalar fields would exist in
Nature, axions
could not only solve the {\em non-baryonic}  dark matter problem
\cite{Tu99}, but their gravitationally confined 
mini--clusters, the {\em axion stars}, would also represent the observed  
MACHOs in our Galaxy.

\section{Outlook: Gravitationally confined Hawking radiation?}
Commonly for the Bekenstein--Hawking radiation the spacetime geometry 
is treated as a given fixed background, e.g. the Schwarzschild solution.
However, due to the universality of gravitational interaction, 
the evaporating quantum field, say a scalar field 
$\hat\Phi$, may have a ``back-reaction" upon the spacetime geometry via the 
semi--classical Einstein equation
\be
   G_{\mu \nu }  =  -\kappa \langle0\vert T_{\mu \nu } 
   (\hat\Phi )\vert 0\rangle \, .\label{back}
\ee  
For instance, a `bouncing shell' model \cite{Step} with  retarded time $u$  
leads to $\langle0\vert T_{uu}\vert0\rangle \rightarrow \kappa^2/48
\pi$, the standard Hawking result. 
The situation becomes, however, much more complicated by the fact
that the vacuum expectation value
$\langle0\vert 0\rangle$ of the energy--momentum tensor $T_{\mu \nu}$, 
for instance defined by the point-splitting 
prescription, is not unique. One ambiguity in 
$\langle0\vert T_{\mu \nu } (\hat\Phi )\vert 0\rangle$ 
is of the type 
$m^2G_{\mu \nu }$, i.e.~linear in the curvature,  and can be readily
absorbed in a redefinition of 
the `bare' gravitational constant
$\kappa$. However, the next order corrections are  quadratic in the 
curvature and 
therefore  of the same one--loop order  arising
from the notorious {\em nonrenormalizability} of perturbative quantum
gravity, cf.~Ref.\cite{Wald}, p.~90. To some extent, the finite part of 
such higher order
curvature counterterms 
in the Lagrangian can be simulated by a self-interaction potential 
$U(\hat\Phi )$, cf.~\cite{BMMO97}.

Already on the semiclassical level 
one could ask the question what happens 
to the (massive) particles associated with the second  quantized field
$\hat\Phi$ in a {\em patch} of some strong gravitational background field? 
Could the  
particles created by the 
{\em Unruh effect} instead of evaporating to infinity rather
form a {\em bound state} within their  {\em self-consistently} 
generated gravitational
field? Moreover, could it be possible that the full back-reaction on the 
geometry is strong enough lead to a curved
spacetime {\em without} horizon and singularities, similarly as in some exact 
solvable (2+1)--dimensional models? Actually
some aspects of this issue were already answered by
Ruffini and Bonazzola \cite{RB69} for a spherically symmetric
self-gravitating configuration of N particles
in a Hartree--Fock approximation. Thus the back--reaction (\ref{back})
may lead us back
exactly to some stable branch of  {\em boson stars}
where the particles are treated on the first quantization level.
These type of stars have an
exponentially decreasing energy density of the scalar field, an
{\em `exosphere'} of particles in
the stable state of equilibrium of particle creation and annihilation.
Moreover, for these
type of compact objects with an effective radius close to the last
stable Kepler orbit an event horizon is suppressed due to
the back-reaction (\ref{back}).

Below the Kaup limit, we have seen that such 
macroscopic quantum states are absolutely stable, at higher central
densities the configuration becomes more and more unstable, and 
undergoes complete gravitational collapse.
 
So could it be that the picture of an evaporating black hole is 
just a first order semi-classical 
approximation; rather, below
some mass limit, we may end up in a self-consistent state of a 
boson or fermion star 
with a {\em gravitationally confined} Hawking radiation, a 
{\em quantum geon}?

\section*{Acknowledgments}
We would like to thank John Barrow, Andrew Liddle,
Alfredo Mac\'{\i}as, and Diego Torres
for useful discussions, literature hints, and support. 
Moreover, we are grateful to 
the Referee for pointing out 
Ref. \cite{FKZ94} and the compatibilty of our  axion mass 
 with the independent estimates  
of Ref. \cite{GV99}.  
This work was partially supported by  CONACyT, grant No. 28339E, and the 
joint German--Mexican project DLR--Conacyt
E130--2924 and MXI 009/98 INF. One of us (E.W.M.)  
thanks Noelia M\'endez C\'ordova for encouragement.
F.E.S.~was supported by an European Union Marie Curie TMR fellowship.
 

\nonfrenchspacing

\end{document}